# Non-LTE ionization potential depression model for warm and hot dense plasma

Chensheng Wu[1,2], Fuyang Zhou[1],* Yong Wu[1], Jun Yan[1], Xiang Gao[1],† and Jianguo Wang[1]

[1] Institute for Applied Physics and Computational Mathematics, Beijing 100088, China

[2] Faculty of Science, Kunming University of Science and Technology, Kunming 650500, China

**Abstracts**

For warm and hot dense plasma (WHDP), the ionization potential depression (IPD) is a key physical parameter in determining its ionization balance, therefore a reliable and universal IPD model is highly required to understand its microscopic material properties and resolve those existing discrepancies between the theoretical and experimental results. However, the weak temperature dependence of the nowadays IPD models prohibits their application through much of the WHDP regime, especially for the non-LTE dense plasma produced by short-pulse laser. In this work, we propose a universal non-LTE IPD model with the contribution of the inelastic atomic processes, and found that three-body recombination and collision ionization processes become important in determining the electron distribution and further affect the IPD in warm and dense plasma. The proposed IPD model is applied to treat the IPD experiments available in warm and hot dense plasmas and excellent agreements are obtained in comparison with those latest experiments of the IPD for Al plasmas with wide-range conditions of 70-700 eV temperature and 0.2-3 times of solid density, as well as a typical non-LTE system of hollow Al ions. We demonstrate that the present IPD model has a significant temperature dependence due to the consideration of the inelastic atomic processes. With the low computational cost and wide range applicability of WHDP, the proposed model is expected to provide a promising tool to study the ionization balance and the atomic processes as well as the related radiation and particle transports properties of a wide range of WHDP.

**Introduction**

WHDP widely exists in stars or giant planets in universe[1, 2], and can be created in experiments with high power lasers[3-6] and Z pinches[7-10]. For atoms embedded in dense plasma, the atomic parameters will be significantly altered due to the complicated many-body interactions with the surrounding plasma[11-18]. IPD is one of the most important phenomena in dense plasma environment. IPD can significantly influence the ionization balance[19, 20] and further impact some important optical and thermodynamic properties of the plasma[21, 22]. Especially, large discrepancy between theory and experiment exists in iron opacity[8, 10], which may attribute to both the flowed experiment and the incomplete or inaccurate approximation in modeling[23]. Therefore, accurate IPD model is of fundamental importance for solving the opacity discrepancy, as well as the studies of astrophysics, inertial confinement fusion and the matter in extreme conditions[24-27].

With recent experimental advances in creating a uniform, well-characterized, and high-energy-density plasmas, it becomes possible to precisely measure the energy level shift as well as IPD of dense plasmas directly[3-5, 28, 29]. It was found that the measured values disagree with predictions of the widely used Stewart-Pyatt (SP)[3, 5, 30] and Ecker-Kröll (EK)[4, 31] analytical models. The experiments performed at the Linac Coherent Light Source (LCLS) have shown that their measurements prefer the earlier EK model[3, 5]. On the other hand, a later experiment at the ORION laser system, which created the plasmas with higher densities and temperatures than those of LCLS, is in better agreement with the SP model[4]. These discrepancies have stimulated extensive theoretical investigations, including two-step average atom model[32], ion-sphere models[20], fundamental plasma calculations[33, 34], classical molecular dynamics[35] and Monte Carlo simulations[36], quantum statistical theory[37], as well as ab-initio calculations based on the density functional theory[38, 39]. Most of these theoretical studies were focused either in the vicinity of zero temperature[38, 39] or under the local thermal equilibrium (LTE) conditions[20, 32-34, 36-39], which were validated by limited experimental data in a rather narrow range of conditions. The applications of these IPD models may be restricted by the large computational cost or limited applicability. Recently, Jin et al developed a Monte Carlo molecular dynamics (MD) method for the transient IPD evolution under nonthermal conditions, where the IPD in each MD time step is calculated by considering the screening effects of the classical free electrons[40]. Nevertheless, the reason underlying the failures of the EK and SP model in different plasma conditions are still not clear. How to construct a unified model for explicitly understanding the fundamental microscopic mechanism of IPD is urgently required.

Generally speaking, the desirable IPD model should capture the substantial variations in atomic dynamical processes with the change of plasma condition over a wide range. For example, in laser-induced plasma in ICF experiments, the temperature, pressure, and density of the plasma often vary several orders of magnitude[26]. With such a wide range of parameter space, the variation in related atomic processes rates such as the three-body recombination (TBR) and collisional ionization (CI) which connect with the density of plasma electrons $\rho_0^2$ and $\rho_0$ respectively, would change drastically[41, 42]. The plasma frequency (in units of Hz) for a given density $\rho_0$ (in units of cm$^{-3}$) could be estimated by the simple expression $f_p = 8.977 \times 10^3 \rho_0^{1/2}$ Hz[43]. The laser-produced dense plasma at a density of $1\times10^{24}$ cm$^{-3}$ is far from being a LTE system with its plasma frequency greater than $9\times10^{15}$ Hz. As the TBR process would change the distribution of bound electrons significantly and results in a prominent modification of the plasma screening potential, considering these microscopic atomic rates for non-LTE processes in the IPD modeling would be indispensable. Especially for modeling the IPD of 1s hollow Al ions observed in LCLS experiments[3], the system is far from LTE since their lifetimes are only about 2-3 fs. We would also like to refer to the two recent line shift experiments under different plasma conditions[28, 29], where different versions of ion-sphere models with the LTE Fermi-Dirac distribution[29, 44, 45] failed to explain the temperature dependence of the two experiments[46]. In contrast, our previous atomic-state-dependent screening model with preliminary consideration of the non-LTE effect give results in good agreement with both experiments[47], which indicates the LTE based methods are inadequate in modelling the WHDP.

In this paper, we propose a non-LTE IPD model for calculating the IPD of ions embedded in WHDP In the model, the contributions of different atomic processes on the IPD can be investigated in a unified framework. The feasibility and validation of the proposed model are demonstrated by reproducing the recent experimental IPD results with wide temperature and density ranges. Moreover, excellent agreements between our predictions and measured IPD of highly-excited hollow Al ions are also achieved, and the IPD of such non-LTE system are found to be highly sensitive to the related specific atomic processes.

**Theory**

For dense plasma, the distribution of the plasma electrons will screen the Coulomb interactions between nucleus and the bound electrons, and then significantly affects the IPD. Therefore, in the first order perturbation, the value of IPD can be expressed as

$$\Delta I = \langle \Psi | V_{scr} - V_{iso} | \Psi \rangle - V_{scr}(r_s), \quad (1)$$

where $|\Psi\rangle$ is the atomic orbital function (AO) of bound electron, $V_{iso}$ is the effective potential felt by an isolated ion, $V_{scr}$ is the screened potential felt by the target ions embedded in plasma and can be obtained by

$$V_{scr}(Q, \boldsymbol{r}) = -\frac{Z}{r} + \int \frac{\rho_b + \delta\rho(\boldsymbol{r}')}{|\boldsymbol{r}-\boldsymbol{r}'|} d\boldsymbol{r}'. \quad (2)$$

Here, $Z$ is the nuclear charge number and $Q$ is the ionization degree, $\delta\rho(\boldsymbol{r})$ is the plasma electron density fluctuation induced by the presence of the target ion, $\rho_b$ is the density distribution of bound electrons of the target ion and calculated by using Multi-configuration Dirac-Fock (MCDF) method[48]. In Eq.(1), the contribution of electron screening on IPD is described by the difference between $V_{scr}$ and $V_{iso}$, and the electron can be ionized in plasma when its energy become higher than $V_{scr}(r_s)$ with $r_s = (3Q/4\pi\rho_0)^{1/3}$ the Wigner–Seitz radius [32]. Therefore, the key issues of IPD calculation are the screened potential $V_{scr}$ and the plasma electron density fluctuation $\delta\rho(\boldsymbol{r})$.

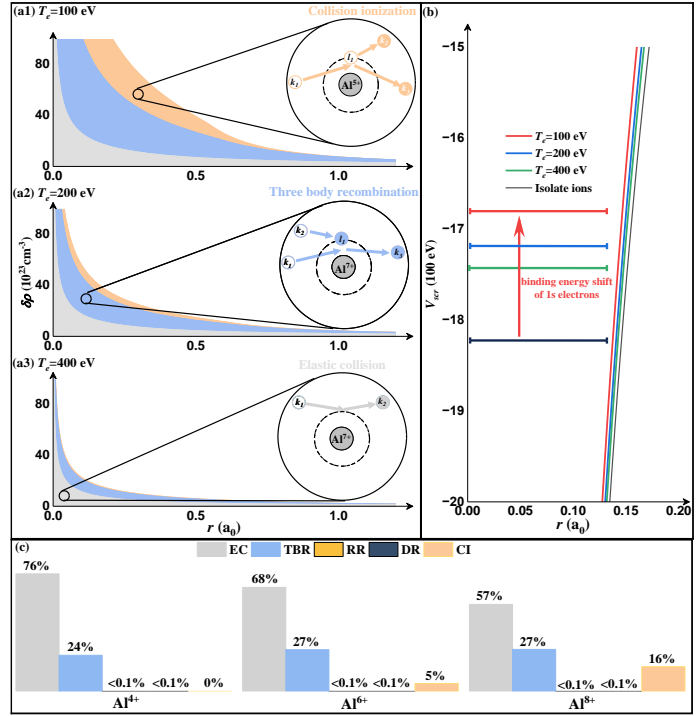

Figure 1. (a) the contribution from different inelastic atomic processes to the plasma electron density fluctuation $\delta\rho$ of $Al^{7+}$ in solid density and $T_e = 100, 200, 400$ eV. (b) the screened potentials and binding energies of 1s electrons of $Al^{7+}$ embedded in the related plasmas. (c) the contribution of different atomic processes to the total IPD in solid density Al plasma with $T_e = 85$ eV. EC: elastic collision, TBR: three-body recombination, RR: radiative recombination, DR: dielectronic recombination, CI: collisional ionization.

Under LTE conditions, the $\delta\rho(r)$ would be determined by the energy structures of the system only, which is irrelevant with the specific atomic processes. For WHDP, the plasma electron distribution should be determined by all atomic processes between electrons and the target ion. In EK, SP and most of previous LTE theoretical models, only free electrons with total energy $E > 0$ are considered, where only the elastic collision (EC) process of electrons can happen. But the plasma electrons can recombine with ion temporarily, and the plasma electrons can also distribute into negative-energy states with $E < 0$ through three-body recombination (TBR), dielectronic recombination (DR) and radiative recombination (RR) processes. For WHDP, it is found that the contributions of inelastic atomic processes on the plasma electron distribution become important as show in figure 1(a). Then, the plasma

electron density fluctuation is expressed as

$$\delta\rho(r) = \frac{1}{\pi^2}\left[\int_{\varepsilon_0}^{-V_{scr}(Q,r)} f(\varepsilon,r)\varepsilon^{1/2}d\varepsilon + \int_{-V_{scr}(r)}^{\infty} f_{FD}(\varepsilon,r)\varepsilon^{1/2}d\varepsilon\right] - \rho_0. \quad (3)$$

Here, $\rho_0$ is the mean electron density, $f_{FD} = 1/[\exp[(\varepsilon + V_{scr} - \mu)/T_e] + 1]$ is the Fermi-Dirac distribution with chemical potential $\mu$ and the temperature of plasma electrons $T_e$. $\varepsilon$ is the kinetic energy of electrons and $\varepsilon_0$ is the lower limit of $\varepsilon$ since the negative-energy electron prefers to populate the outer orbital unoccupied. When the rates of EC process is much larger than inelastic atomic processes, the distributions of free electrons with total energy $E = \varepsilon + V_{scr}(r) > 0$ are assumed to be in equilibrium and can be described by $f_{FD}$. $f(\varepsilon, r)$ is the distribution of negative-energy electrons, which can be determined theoretically by solving the rate equations with considering all inelastic atomic processes of the relevant ions with different charge states. For simplicity, the contribution of TBR processes can be approximated to analytical form as

$$f_{TBR}(\varepsilon, r) = 2\sqrt{\frac{E_r}{\pi T_e}} \frac{e^{-\frac{E_r}{T_e}}}{Erf(\sqrt{\frac{E_r}{T_e}})} f_{FD}(\varepsilon, r), \quad (4)$$

by employing steady-state approximation and the classical scattering rates[47]. It is only dependent on the distribution of free electron and the state of recombined ions, and thus the complicated calculations of rate equations can be greatly simplified. Here, $E_r = |\varepsilon + V_{scr}(Q, r)|$ is the absolute value of the the total energy, $Erf(x)$ is the error function. For WHDP, TBR processes dominate the distribution of negative-energy electrons and the contributions from different inelastic atomic processes to IPD, as shown in figure 1. Therefore, by applying the first-order perturbation approximation, the contributions from other inelastic atomic processes can be evaluated by comparing its rate with TBR as

$$\eta_P = \frac{R_P^{Q+} N_P^{Q+}}{R_{TBR}^{q+} N_{TBR}^{q+}}, \quad (5)$$

where $R_P^{Q+}$ and $N_P^{Q+}$ are the rate coefficients and the populations of the initial state ions of inelastic atomic processes $P$, $R_{TBR}^{q+}$ and $N_{TBR}^{q+}$ are the ones of TBR process. The negative-energy electron distribution $f(\varepsilon, r)$ can be calculated as $f = f_{TBR} +$

$\sum_P f_P$, in which the contribution of a specific process $P$ is $f_P = \eta_P f_{TBR}$. Then, the contributions from different atomic processes can be included approximately in a unified framework.

**Results and Discussion**

As the first application, the present model is used to treat the problem of IPD in the latest LCLS experiments[3, 5], in which the density of ions $\rho_i$ is close to the one of solid material, and the electron temperature $T_e$ is estimated to range from 70 to 180 eV[3]. As discussed above, the target ion $A^{q+}$ can combine with a negative-energy electron to $A^{*(q-1)+}$ temporarily though TBR, RR and DR processes. Moreover, $A^{*(q-1)+}$ can also be produced by the electron-ion collisional ionization (CI) processes of $A^{(q-2)+}$. Using the treatment described in Supplemental Material[49], the rate coefficients for different inelastic processes and their contribution to IPD can be calculated. In Supplemental Material[49], the rate coefficients of different inelastic atomic processes and their contributions to IPD are presented for Al plasma with solid density and temperature in 85eV. Based on the rate coefficients, the contribution of elastic and inelastic processes to IPD in $Al^{4+}$, $Al^{6+}$ and $Al^{8+}$ with solid density and $T_e = 85\ eV$ are calculated and presented in figure 1(c). It is found that contribution of TBR and CI processes on IPD is important for the present case, especially for the ions in high charge states, and the ones of DR and RR processes are much smaller. Therefore, beside the elastic collision processes, TBR and CI processes should be considered in the IPD modelling.

For validation of the present model, IPDs of different charge states of Mg and Al ions for different plasma temperatures are calculated and are compared with the latest experimental data in figure 2 (a) and (b). According to the experimental conditions, $T_e = 70, 85$ and 100 eV with $\rho_i$ in solid material are applied in the calculation. It is found that the present IPD results $\Delta I$ agree well with the experimental values. Being similar to the previous LTE models, the results $\Delta I_0$ of elastic collision (EC) model are calculated without including the contribution from inelastic atomic processes and found to significantly underestimate the IPDs. It can be concluded that contribution of inelastic atomic processes on IPD is important for the present case of warm and dense plasma.

In detail of figure 2 (a), the IPDs of Mg ions with $T_e = 70$ eV are in good agreement with experimental data in most of the ionization degrees, but higher than experimental results in $Mg^{9+}$. With higher temperature applied ($T_e = 85$ and 100 eV), good agreement is achieved between the calculation and experiment for $Mg^{9+}$. Similar phenomenon also exists for Al ions. These discrepancies at high ionization degrees between our calculations with $T_e = 70$ eV and experiment may originate from the underestimation of the plasma temperature. In experiments, pump photon energies need to be higher for producing higher ionization degrees, and thus it is very reasonable to assume a higher $T_e$ for ions with higher ionization degree[3, 5]. After introducing a higher temperature, the theoretical results become consistent with the experimental data. On the other hand, without including the contribution of inelastic atomic processes, the calculated IPDs $\Delta I_0$ are nearly the same for different temperatures. In the present model, a stronger temperature dependence is induced by considering inelastic atomic processes, which are sensitive to the temperature and density of plasma.

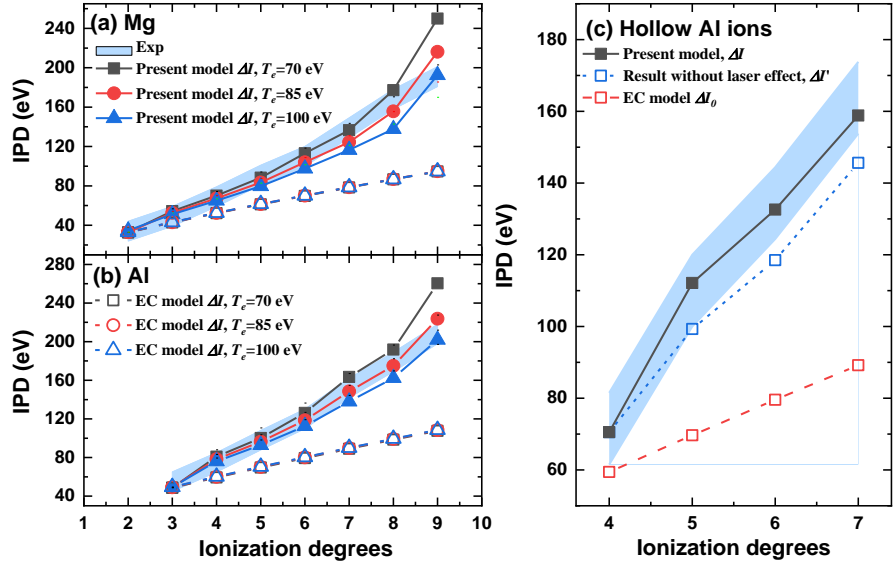

Figure 2 The IPD calculated in present model and compare with the LCLS experiment[3]. (a) (b) the IPD of Mg and Al ions with solid density and $T_e = 70, 85$ and $100$ eV. $\Delta I$ represent the results of present model, and $\Delta I_0$ are the results of EC model without the contribution from inelastic atomic processes. (c) the IPD of hollow Al ions with solid density and $T_e = 70$ eV. For such non-LTE system, the results are sensitive to the specific atomic processes, and the results $\Delta I'$ without considering the influence of photoionization due to the pump laser are also presented.

Furthermore, the present non-LTE IPD model is applied to calculate the K-edges of *1s* hollow Al ions which are also observed in the LCLS experiment[3]. The *1s* hollow Al ions decay very fast through radiative transition and Auger processes, and thus their lifetimes are only 2-3 fs, which is even smaller than the plasma equilibrium time and the timescale of the inelastic atomic processes. Therefore, the relaxation of plasma electron distributions should be neglected in the short lifetime of *1s* hollow ions, and the plasma electron distributions around the hollow ions can be obtained from the ones before the ionization of *1s* electron. On the other hand, for such non-LTE system, the effect of the pump laser on the IPD becomes important. In the experiments, the *1s* hollow Al ions are produced by photoionization due to the pump laser. The selection effect of pump laser on the negative-energy electron distributions should be considered, due to that the photoionization cross sections are different between ions with and without the presence of negative-energy electron. Considering this influence, the proportion of the hollow ions with and without negative-energy electron can be expressed as

$$\frac{N_{1s}^{*(q-1)+}}{N_{1s}^{q+}} = \frac{N^{*(q-2)+}\sigma^{*(q-2)+}(\omega)}{N^{(q-1)+}\sigma^{(q-1)+}(\omega)}. \quad (6)$$

Here $N_{1s}^{*(q-1)+}$ and $N_{1s}^{q+}$ are the population of *1s* hollow ions with and without combining negative-energy electron, respectively. $\omega$ the photon energy of pump laser, $\sigma^{*(q-2)+}(\omega)$ and $\sigma^{(q-1)+}(\omega)$ are the photoionization cross-section of the related ground states ions $A^{*(q-2)+}$ and $A^{(q-1)+}$, respectively. $N^{*(q-2)+}$ and $N^{(q-1)+}$ are the population of related ground states ions. Then the distribution of negative-energy electrons in 1s hollow ions $f^{1s}(\varepsilon, r)$ can be expressed as $f^{1s} = \eta_\sigma f$ with $\eta_\sigma(\omega) = \sigma^{*(q-2)+}(\omega)/\sigma^{(q-1)+}(\omega)$.

Based on the above treatments, the IPD of 1s hollow Al ions are calculated with solid density and $T_e = 70$ eV and compared with experiment in figure 2(c). The uncertainty of measurement is given as 10 eV, which is estimated by IPD of unionized ions in Ref. [3]. The present results $\Delta I$ are in great agreements with experimental measurements, which further verify the reliability of present model on such non-LTE system. For comparison, the results of EC model $\Delta I_0$ are presented and are found to significantly underestimate the IPD of hollow Al ions, which elucidate the importance of inelastic collision processes. Furthermore, the results $\Delta I'$ without considering the influence of photoionization due to the pump laser are also presented in figure 2(c), and are found to underestimate the IPD. It indicates that the non-LTE electron distribution are sensitive to the specific atomic processes, and an insufficient consideration of the corresponding atomic processes would lead to an evident divergence of the prediction of non-LTE IPD.

In order to have a deep insight into the temperature dependence of IPD, figure 3 illustrates the present results of IPDs of ground-state $Al^{7+}$, $Al^{9+}$ and $Al^{11+}$ ions as a function of temperature, the results of widely applied SP, EK, Ion-Sphere (IS) and Debye-Hückel (DH) models[50] and experimental data from LCLS and Orion[3-5] are also shown. It is found from figure 3 that in the temperature area with $T_e \sim 100$ eV and electron density in $5.0 \times 10^{23} \text{cm}^{-3}$, which is related to the condition of the LCLS experiment[3, 5], the present results are close to EK model and experiment, while SP model is found to underestimate IPD evidently[3, 5]. For higher temperature and density with $T_e > 550\ eV$ and electron density in $1.5 \times 10^{24} \text{cm}^{-3}$, which is related to the condition of Orion experiment[4], our results are evidently lower than EK model and are close to SP model. In such plasma condition, EK model have been found to overestimate the IPD and our model has better agreement with measurement[4]. For high temperatures, the free electrons dominate the distribution of plasma electrons as shown in figure 1, and thus our model will converge to the LTE SP model. In the middle temperature area with $250 < T_e < 375\ eV$, the feasibility of our model is also validated by the measured line shifts of $K\alpha$ line in $Al^{11+}$, which is the difference of binding energy shifts between 1s and 2p due to the plasma environment. Excellent agreement between our model and experiment has been demonstrated in our previous work[47], while IS models underestimate the line shifts[45]. The above results manifest that the present model exhibits a stronger temperature dependence and can be applied to treat plasma with broader range of temperatures and densities.

For further illustration, the results without the contribution of inelastic atomic processes are obtained and also presented in figure 3. It is found that the consideration of inelastic atomic processes leads to the stronger temperature dependence. But in SP model, only the contribution of free electrons on screening effect is included, and thus its variation tendency of temperature is close to the present model without inelastic atomic processes. On the other hand, the EK model is only related to the plasma density and thus has no explicit temperature dependence at all[3]. In a word, it is the inadequate description of the temperature dependence that leads to the SP and EK model not work in warm-dense plasmas and high temperature ranges, respectively.

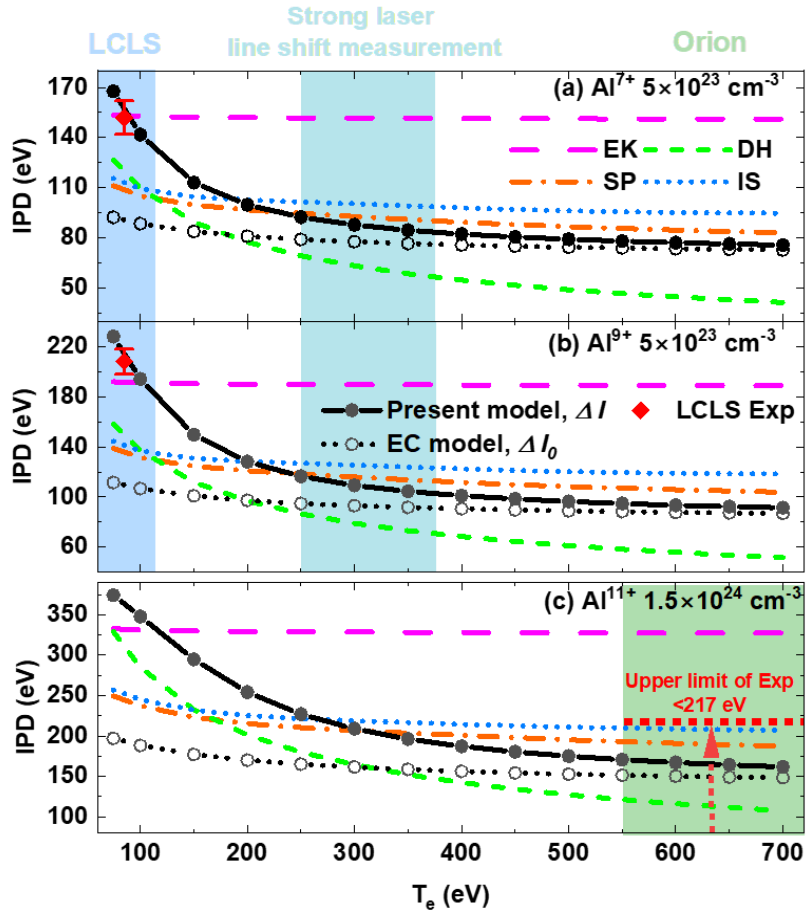

Figure 3 The IPD of $Al^{7+}$, $Al^{9+}$ and $Al^{11+}$ ions as a function of temperature (from 75 to 700 eV) at a fixed electron density of $5.0 \times 10^{23} cm^{-3}$ for $Al^{7+}$ and $Al^{9+}$, and $1.5 \times 10^{24} cm^{-3}$ for $Al^{11+}$, respectively. $\Delta I$ represent the results of present model, and $\Delta I_0$ are the results of EC model without the contribution of inelastic atomic processes. The results from other theoretical model and experiment are also shown for comparison.

**Conclusion**

In conclusion, a non-LTE IPD model is proposed to include the contributions of different inelastic atomic processes. For non-LTE system, the results are expected to be sensitive to the specific atomic processes. With appropriate considering the characteristic atomic processes involved in different atomic systems, our model had been successfully applied to WHDP with distinctly different conditions. Instead of directly solving the complicated rate equations, the contributions of different inelastic atomic processes on the electron distribution are approximately obtained based on the analytical form of contribution of TBR process, and thus the computational complexity of our model is nearly same to the LTE model. Although we have made reasonable simplification in solving the rate equations in our model to increase the usability in practical applications, the validation of the model has been justified by well reproducing the IPD in the latest warm and dense plasmas experiments. In particular, an excellent agreement is achieved in comparison with the IPD of hollow Al ions for the first time.It is found that the inelastic atomic processes are important in modeling the IPD of WHDP, especially for the TBR and CI processes. A stronger temperature dependence of IPD is demonstrated by including the inelastic atomic processes, and the inadequate consideration of the atomic processes make the SP and EK model only valid for plasma with limited temperature and density conditions. And it is exactly the motivation of present work to provide a promising tool to treat the IPD for the plasmas with a wide range of temperatures and densities, which will benefit the future studies of the radiation and particle transports properties, and help to resolve those existing challenging problems, such as the discrepancies between measured and modeled iron opacity.